\documentclass[prb,twocolumn,showpacs,floatfix,amsmath,amssymb,superscriptaddress]{revtex4-2}
\usepackage{amsfonts}
\usepackage{stmaryrd}
\usepackage{bbm}
\usepackage{mathrsfs}
\usepackage{tipa}
\usepackage{amssymb}
\usepackage{txfonts}
\usepackage{graphicx}
\usepackage{dcolumn}
\usepackage{epstopdf}
\usepackage[colorlinks,linkcolor=blue,urlcolor=blue,citecolor=blue]{hyperref}
\usepackage{multirow}
\usepackage{subfigure}
\usepackage{url}
\usepackage{upgreek}
\usepackage[utf8]{inputenc}
\usepackage[english]{babel}
\usepackage{siunitx}

\begin{document}

\title{Field and Polarization Dependent Quantum Spin Dynamics in Honeycomb Magnet Na$_2$Co$_2$TeO$_6$: Magnetic Excitations and Continuum}

\author{Patrick~Pilch}
\affiliation{Department of Physics, TU Dortmund University, 44227 Dortmund, Germany}

\author{Laur~Peedu}
\affiliation{National Institute of Chemical Physics and Biophysics, 12618 Tallinn, Estonia}

\author{Anup~Kumar~Bera}
\email{akbera@barc.gov.in}
\author{S.~M.~Yusuf}
\affiliation{Solid State Physics Division, Bhabha Atomic Research Centre, Mumbai 400 085, India}
\affiliation{Homi Bhabha National Institute, Anushaktinagar, Mumbai 400 094, India}

\author{Urmas~Nagel}
\author{Toomas~Rõõm}
\affiliation{National Institute of Chemical Physics and Biophysics, 12618 Tallinn, Estonia}

\author{Zhe Wang}
\email{zhe.wang@tu-dortmund.de}
\affiliation{Department of Physics, TU Dortmund University, 44227 Dortmund, Germany}

\date{\today}

\begin{abstract}
We report terahertz spectroscopic measurements of quantum spin dynamics in the spin-1/2 honeycomb magnet Na$_2$Co$_2$TeO$_6$ as a function of applied magnetic field with different terahertz polarizations.
Distinct field dependencies of the resolved spin dynamics are identified in three regimes, which are separated by two critical fields at $B_{c1}\approx 7$ and $B_{c2}\approx 10$~T.
A polarization selective continuum is observed in the intermediate phase, featuring spin fluctuations of a proximate quantum spin liquid.
\end{abstract}

\maketitle

The spin-1/2 honeycomb magnet with bond-dependent nearest-neighbor Ising exchange interaction -- the so-called Kitaev interaction -- is a remarkable example among the few exactly solvable models of two dimensions \cite{kitaev2006anyons}.
This system hosts an exotic ground state of quantum spin liquid, which without long-range magnetic order even down to the lowest temperature is a long-sought unconventional state of matter \cite{anderson1973resonating}.
Featured by long-range quantum entanglement and fractionalized excitations \cite{savary2016quantum,wen2019experimental,Broholm20}, the Kitaev quantum spin liquid exhibits various unconventional properties \cite{Hermanns18,Takagi2019}, such as half-integer quantization of thermal Hall conductivity \cite{kitaev2006anyons,Nasu17} and magnetic continua \cite{Knolle14,Winter2017,Hickey2019}.

The realization of the strong bond-dependent exchange anisotropy is not straightforward in a solid-state material.
Transition metal elements with strong spin-orbit coupling, such as 5\textit{d} Ir$^{4+}$ and 4\textit{d} Ru$^{3+}$ ions, in a proper crystal-field environment were proposed \cite{jackeli2009mott} and indeed found in iridates \cite{Rau16} and $\alpha$-RuCl$_3$ to contribute a significant Kitaev interaction.
This has been evidenced by the observation of magnetic continua (see e.g. \cite{Banerjee17,Do2017,Banerjee2018,WangPRLMagn,
Sahasrabudhe20,Wulferding2020}) and peculiar thermal Hall effects \cite{Yokoi21,Bruin2022,Czajka2023}. 

Recent theoretical analysis suggested that based on the 3$d^7$ Co$^{\mathrm{2+}}$ ions with a high-spin $t^5_{2g}e^2_{g}$ configuration, a material realization of the Kitaev honeycomb model should also be possible \cite{LiuPRBPseudospin, SanoPRBkitaev}. Various candidate compounds have been studied \cite{kim2021spin}, among which BaCo$_2$(AsO$_4$)$_2$ \cite{zhong2020weak} and Na$_2$Co$_2$TeO$_6$ \cite{lefranccois2016magnetic,bera2017zigzag} are particularly promising. Their long-range magnetic orders can be suppressed by applying an in-plane magnetic field \cite{zhong2020weak,yao2020ferrimagnetism,lin2021field}, similar to the field dependent behavior of $\alpha$-RuCl$_3$.
However, the role of Kitaev type interaction in these compounds remains elusive.
On the one hand, in addition to magnonic excitations \cite{Shi21}, a broad magnetic continuum was observed in BaCo$_2$(AsO$_4$)$_2$ \cite{Zhang2023} by terahertz spectroscopy corresponding to the $\Gamma$ point, which signals a quantum spin liquid.
On the other hand, the inelastic neutron scattering spectra at finite momentum transfers \cite{Halloran23} are better described by an extended XXZ model with Heisenberg type exchange interactions \cite{Das21}.
For Na$_2$Co$_2$TeO$_6$ the knowledge of low-energy magnetic excitations at zero magnetic field is not sufficient to settle a spin Hamiltonian with dominant Kitaev interaction against a Heisenberg XXZ model \cite{songvilay2020kitaev,kim2021antiferromagnetic,lefranccois2016magnetic,mao2020theoretical, kim2021antiferromagnetic,samarakoon2021static,lin2021field,sanders2022dominant,yao2022excitations}. 

The long-range antiferromagnetic order stabilized below the Néel temperature of $T_N \approx 26$~K in Na$_2$Co$_2$TeO$_6$ can be suppressed by a relatively large in-plane magnetic field $B_{c2}\approx 10$ -- 12~T, above which the system enters a spin-gapped state \cite{lefranccois2016magnetic,bera2018magnetic, bera2017zigzag, berthelot2012studies, chaudhary2018evidence, lin2021field,hong2021strongly, mukherjee2022ferroelectric,murtaza2021magnetic, samarakoon2021static,xiao2019crystal, yang2022significant, yao2020ferrimagnetism,lee2021multistage,xiao2021magnetic,takeda2022planar}.
Studies by using a variety of experimental techniques have revealed at least one lower critical field $B_{c1}\approx 6$ -- 8~T, thereby determined an intermediate phase.
While phenomenologically this is very similar to the field dependent behavior in $\alpha$-RuCl$_3$, the nature of the intermediate phase in Na$_2$Co$_2$TeO$_6$ is still under debate.

Motivated by these results, we perform high resolution terahertz spectroscopy as a function of applied magnetic field in the \textit{ab} plane to resolve the spin dynamics of Na$_2$Co$_2$TeO$_6$.
Three phases are identified by their distinct field dependence of characteristic spin dynamics. In particular, an intermediate phase between 7 and 10~T is featured by a continuum, which is absent in the lower- and higher-field phases, providing evidence for a possibly dominant Kitaev interaction. 

High-quality single crystals of Na$_2$Co$_2$TeO$_6$ were synthesized by a self-flux method \cite{BeraUnP}, with the initial polycrystalline samples prepared by the solid state method \cite{bera2017zigzag}.
Terahertz (THz) transmission experiment was performed on the single crystals with the propagation direction perpendicular to the crystallographic \textit{ab} plane by using a Sciencetech SPS200 Martin-Puplett type spectrometer \cite{Amelin22}.
Transmission spectra were measured above and below the Néel temperature in zero field, and at 3~K in external magnetic fields applied in the \textit{ab} plane up to 17~T by using a magneto-optic cryostat equipped with a superconducting magnet.
A rotatable polarizer was set in front of the sample to tune the THz polarization.
To enhance the sample absorption, two pieces of single crystals with a typical size of $3 \times 3$~mm$^2$ in the \textit{ab} plane were aligned and stacked together with a thickness of about 0.5~mm.

\begin{figure}[t!]
\centering
\includegraphics[width=0.85\linewidth]{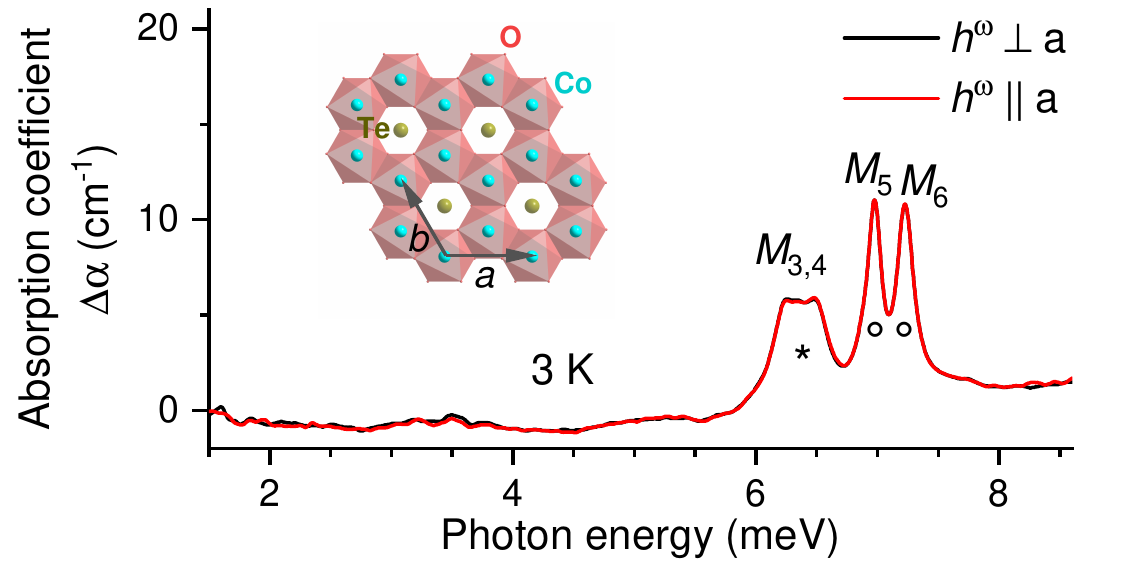}
\caption{THz absorption spectra of Na$_2$Co$_2$TeO$_6$ at 3~K for two polarizations $h^{\omega} \perp a$ and $h^{\omega} \parallel a$. Magnetic excitations are marked by asterisk ($M_{3,4}$) and circles ($M_5$ and $M_6$). Inset: Honeycomb lattice of CoO$_6$ octahedra in the crystallographic \textit{ab} plane.}
\label{fig1:zero}
\end{figure}

We determine the magnetic excitations below $T_N\approx 26$~K by measuring transmitted THz spectra at 3~K in the magnetically ordered phase and also at 30~K slightly above $T_N$ as a reference.
Figure~\ref{fig1:zero} shows the obtained absorption coefficient at zero field for the polarized THz magnetic field $h^{\omega}$ perpendicular and parallel to the crystallographic \textit{a} axis, i.e. $h^{\omega} \perp a$ and $h^{\omega} \parallel a$, respectively.
A spectral range up to 8.6~meV is presented, because the higher energy response is governed by strong phonon absorption.
The spectra are almost identical for the two polarizations, which both exhibit two sharp absorption peaks at 6.98 and 7.23~meV, respectively, as marked by the circles ($M_5$ and $M_6$), and one broader absorption band centered at 6.38~meV, as indicated by the asterisk ($M_{3,4}$).
Since these features disappear above $T_N$, we can assign them as magnetic excitations of the ordered phase in Na$_2$Co$_2$TeO$_6$. 
A momentum transfer by the THz photons is negligible in the experiment, therefore these excitations correspond to the $\Gamma$ point in the reciprocal space.
In the same energy range, inelastic neutron scattering experiment on powder \cite{lin2021field,kim2021antiferromagnetic} and signal crystal samples \cite{yao2022excitations} revealed two excitation bands below $T_N$.
Although these bands were resolved at finite momentum transfers, they are nearly dispersionless thus consistent with our observed excitation modes.   
With higher energy resolution than the neutron experiment, we are able to resolve more excitations whose energies are very close to each other.

\begin{figure}[t]
\centering
\includegraphics[width=1\linewidth]{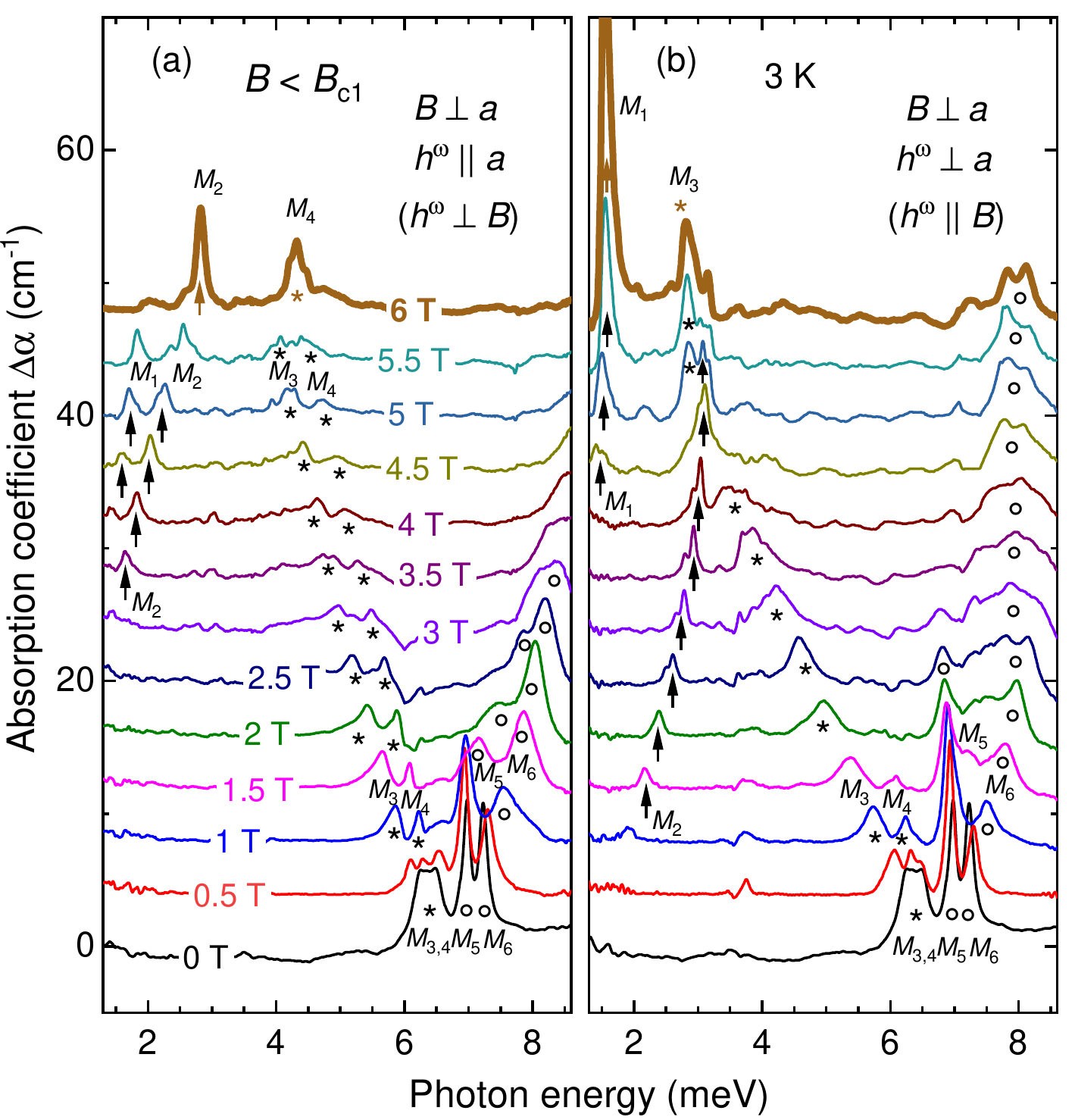}
\caption{Magnetic excitations observed at 3~K in external magnetic fields below a lower critical value $B<B_{c1}\approx 7$~T, which are applied  perpendicular to the crystallographic \textit{a} axis, i.e. $B \perp a$, for two cross polarizations of THz radiation (a) $h^\omega \parallel a$ and (b) $h^\omega \perp a$. The higher-field spectra are shifted upward by constants which are proportional to the field strengths.}
\label{fig2:Bc1}
\end{figure}

\begin{figure}[t]
\centering
\includegraphics[width=1\linewidth]{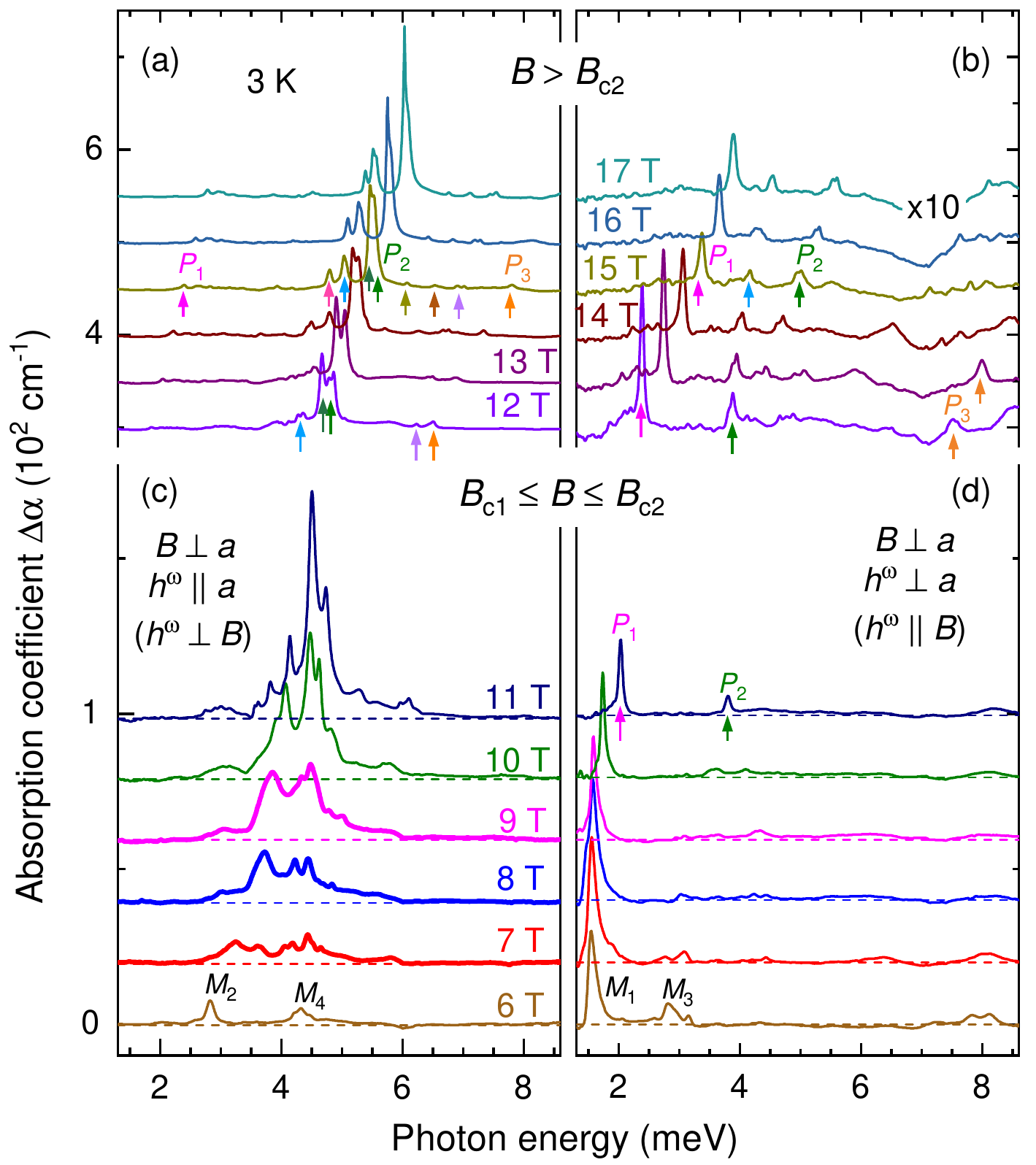}
\caption{Evolution of magnetic excitations at 3~K (a)(b) in the high field regime $B>B_{c2}\approx 10$~T and (c)(d) in the intermediate field range $B_{c1}\leq B \leq B_{c2}$ for configurations (a)(c) $h^\omega \parallel a \perp B$ and (b)(d) $h^\omega \parallel B \perp a$. The spectra in (b) are magnified by a factor of 10 to highlight the relatively weak but clearly resolved modes. The arrows indicate the magnetic excitations, whose eigenfrequencies and linewidths as a function of field are presented in Fig.~\ref{fig4:contour}(a)(b). The spectra are vertically shifted for clarity.}
\label{fig3:Bc2}
\end{figure}

By applying an external magnetic field along the honeycomb bond direction in the crystallographic \textit{ab} plane (see Fig.~\ref{fig1:zero}), i.e. $B \perp a$, we trace the field-dependent evolution of the magnetic excitations in detail.
The obtained spectra for fields $B < B_{c1} \approx 7$~T are presented in Figs.~\ref{fig2:Bc1}(a) and \ref{fig2:Bc1}(b) for $h^{\omega} \parallel a$ and $h^{\omega} \perp a$, respectively, while for $B \geq B_{c1}$ the spectra are shown in Fig.~\ref{fig3:Bc2}.

In contrast to the zero-field data, the spectra of the two polarizations in the applied fields are no more identical but exhibit various similarities mainly below $B_{c1}$ and, more importantly, distinct differences above $B_{c1}$. 
Starting from zero field the absorption band $M_{3,4}$ splits into two absorption peaks $M_3$ and $M_4$ which soften with increasing fields, whereas the $M_5$ and $M_6$ modes shift to higher energies.
Above 4~T two more lower-lying modes are resolved, as marked by arrows ($M_1$ and $M_2$), both of which evolve from the lower-boundary of the low-lying spin-wave excitation \cite{lin2021field} and harden with increasing fields. These are the common features of field dependency shared by both polarizations. In the following we highlight the interesting quantitative differences.

First, the $M_6$ mode hardens more evidently for $h^{\omega}\perp B$ [Figs.~\ref{fig2:Bc1}(a)] and shifts out of our resolvable spectral range already above 3.5~T. In contrast, for $h^{\omega}\parallel B$ [Figs.~\ref{fig2:Bc1}(b)] the initial increase in energy is taken over by a nearly field independent evolution above 2~T. The $M_5$ mode exhibits a similar contrast for the two polarizations, but becomes very weak above 3~T.
Second, different from the contrast of the $M_6$ mode, the softening of the $M_3$ and $M_4$ modes is less evident for $h^{\omega}\perp B$ than for $h^{\omega}\parallel B$.
For $h^{\omega} \perp B$ the $M_4$ mode evolves clearly together with the $M_3$ mode until 5.5~T [Figs.~\ref{fig2:Bc1}(a)], whereas for $h^{\omega}\parallel B$ the $M_4$ mode is very weak already above 2~T.
Third, the eigenenergy of the $M_2$ mode for $h^{\omega}\perp B$ is evidently smaller than for $h^{\omega}\parallel B$, while the peak frequency of the $M_1$ mode is nearly the same for the two polarizations.
Despite these differences, for simplicity we will still use the same nomenclature in the following with the polarization specified.

A rather abrupt change induced by the applied magnetic fields occurs at 6~T. While at 5.5~T the low-energy spectra are characterized by four excitation modes ($M_1$ to $M_4$), at 6~T the spectra are  dominated by only two of them, the peak values of which are greatly enhanced.
Also the THz polarization plays an important role.
While for $h^{\omega}\perp B$ the $M_2$ and $M_4$ modes survive with enhanced intensity above 5.5~T [Fig.~\ref{fig2:Bc1}(a)], for $h^{\omega}\parallel B$ the $M_1$ and $M_3$ modes dominate the low-energy dynamical response. 
These behaviors are reminiscent of the dynamic characteristics of field-induced spin reorientations that were observed in $\alpha$-RuCl$_3$ \cite{Wu18}. 
Nonetheless, a decisive claim of the magnetic order still requires a detailed theoretical analysis of the field dependent spin dynamics \cite{Wu18,krueger2022tripleq}.

The most intriguing dynamic feature appears when the applied magnetic field is further increased, as shown in Fig.~\ref{fig3:Bc2}. Corresponding to the transverse polarization $h^{\omega} \perp B$ [Fig.~\ref{fig3:Bc2}(c)], the 7~T spectrum is dominated by a broad continuum spanning from 2.5 to about 6~meV. Small peaklike features are discernible on top of this continuum, whose line shape is however not well defined.
With increasing fields the continuum evolves further and become strongest at 9~T, on top of which two broad peaks have developed. From 10~T onward the absorption peaks are getting sharper, while the underlying continuum is less evident.
Above 12~T the continuum essentially disappears and the spectra are finally characterized by nine sharp absorption peaks, as indicated by the arrows at 15~T [Fig.~\ref{fig3:Bc2}(a)].
The peaks at higher energies appear to be rather weak from the representation of Fig.~\ref{fig3:Bc2}(a), but in fact can be unambiguously identified by their field dependence. This is clearly seen if the spectra are zoomed by a magnification factor of 10 [see Fig.~\ref{fig3:Bc2}(b)].
The appearance and disappearance of the continuum provide a spectroscopic determination of the field-induced quantum phase transitions, i.e. at $B_{c1}\approx 7$~T and $B_{c2}\approx 10$~T.
These values are consistent with the results of thermodynamic measurements such as specific heat, magnetization \cite{lin2021field}, and thermal conductivity \cite{hong2021strongly,takeda2022planar,Guang23}.

For the longitudinal polarization $h^{\omega}\parallel B$ [see Fig.~\ref{fig3:Bc2}(b)(d)], the high-field excitation spectra exhibit distinctly different characteristics than for the transverse polarization. In particular, an extended continuumlike feature is absent. 
As shown in Fig.~\ref{fig3:Bc2}(d), the sharp $M_1$ mode at 6~T becomes even stronger and sharper at higher fields, while its eigenenergy is nearly constant up to 9~T. 
Above $B_{c2}\approx 10$~T this is followed by a shift of its peak position towards higher energy and a concomitant decrease of its peak intensity until the highest field of 17~T [see Fig.~\ref{fig3:Bc2}(b)].
At 12~T three magnetic excitations are discernible, which are labelled as $P_1$, $P_2$, and $P_3$, as indicated by the arrows in Fig.~\ref{fig3:Bc2}(b).
While $P_1$ evolves from the $M_1$ mode, the other two modes emerge above $B_{c2}=10$~T and harden in higher fields, which are dynamical characteristics for the high-field phase [Fig.~\ref{fig3:Bc2}(c)]. In addition, another mode splits off from the $P_2$ mode (see cyan arrow at 15~T) which exhibits a weaker field dependence.  

\begin{figure*}[t]
\centering
\includegraphics[width=1\linewidth]{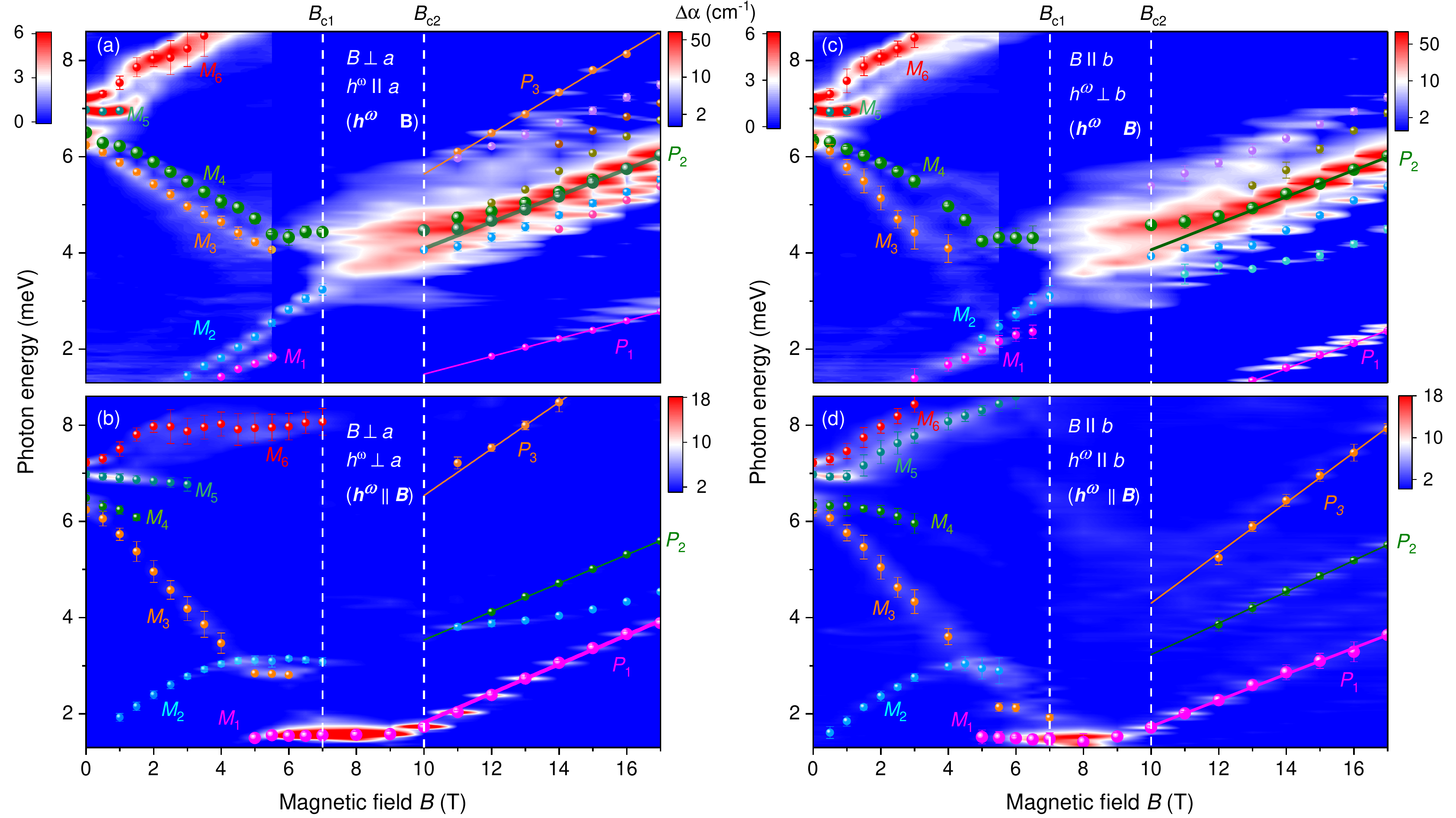}
\caption{Absorption coefficient $\Delta \alpha$ corresponding to magnetic excitations and continuum for the configurations (a) $B \perp h^\omega \parallel a$ and (b) $B \parallel h^\omega \perp a$,
and for (c) $h^\omega \perp B \parallel b$ and (d) $h^\omega \parallel B \parallel b$.
Three field regimes $B<B_{c1}$, $B_{c1}\leq B \leq B_{c2}$, and $B>B_{c2}$ are characterized by distinctly different features of quantum spin dynamics, which are separated by the dashed lines. 
The scatters and bars represent the resonance frequency and linewidth of each magnetic excitations. 
In (a)(c), for fields below and above 5.5~T different scales of color coding are utilized for clarify.}
\label{fig4:contour}
\end{figure*}

These results are summarized in Fig.~\ref{fig4:contour}(a) and \ref{fig4:contour}(b) by a contour plot of the obtained absorption coefficients for the two polarizations $h^\omega \parallel a \perp B$ and $h^\omega \parallel B \perp a$, respectively.
The eigenenergies and linewidths of the well-defined magnetic excitations (see arrows, asterisks, and circles in Fig.~\ref{fig2:Bc1} and Fig.~\ref{fig3:Bc2}) are denoted by scatters and bars, respectively.
In this representation one can readily identify three regimes by their different field-dependent dynamical characteristics. 

While below $B_{c1}$ the dynamics probed via both polarization channels is characterized by sharp magnetic excitation modes, a significant contrast appears above $B_{c1}$.
For the transverse configuration $h^\omega \perp B$, an extended magnetic continuum exists at $B_{c1}\leq B \leq B_{c2}$, which provides possible evidence for fractionalized excitations of a quantum spin liquid, as expected for dominant Kitaev-type interaction in a honeycomb lattice.
In contrast, in the channel of $h^\omega \parallel B$, only a sharp absorption $M_1$ mode is observed in the same field range.
It is worth noting that a sharp absorption profile at the $\Gamma$ point is not an exclusive signature for a long-range magnetic order, but can also be a feature of a quantum spin liquid due to Kitaev-type interaction \cite{Knolle14,Knolle15} even in presence of an external field \cite{Hickey2019}.
In particular, despite the observation that the magnetic order is being partially suppressed in Na$_2$Co$_2$TeO$_6$ \cite{yao2022magnetic}, the peak intensity of the $M_1$ mode still increases above $B_{c1}$.
Therefore, the observed coexistence of magnetic excitation and continuum, although in different polarization channels, points to the importance of Kitaev interaction in the quantum spin dynamics of Na$_2$Co$_2$TeO$_6$.

While the quantum spin fluctuations are enhanced above $B_{c1}$ leading to the emergence of the dominant magnetic continuum, the system may still maintain a long-range magnetic order by properly accommodating the Zeeman energy, especially with the assistance of inter-layer couplings \cite{yao2022magnetic}.
This magnetic order is unlikely the same as the one at zero field, because in contrast to the various magnetic excitations with relatively weak intensity and evident field dependencies below $B_{c1}$, the magnetic excitations for $B_{c1}\leq B \leq B_{c2}$ are sharper and nearly field independent [c.f. also Fig.~\ref{fig2:Bc1}(b) and \ref{fig3:Bc2}(d)]. 

The second field-induced phase transition at $B_{c2} = 10$~T is characterized by the appearance of various magnetic excitations which all harden with increasing field. For $h^\omega \perp B$ [see Fig.~\ref{fig4:contour}(a)] one can spot a new mode at the lowest energy and another one at the highest energy within our spectral range, which are denoted by $P_1$ and $P_3$, respectively [see also Fig.~\ref{fig3:Bc2}(a)]. At the same time, the continuum evolves above $B_{c2}$ into a series of magnetic excitations, of which we label the strongest one by $P_2$. The counterparts of these three representative modes for the other polarization $h^\omega \parallel B$ are correspondingly labelled in Fig.~\ref{fig4:contour}(b).

For both polarizations these three modes exhibit a linear dependence of their eigenfrequencies on the applied magnetic field. A fit by the linear dependence $\hbar\omega = g\mu_B B\Delta S$ corresponding to a magnetic-dipole excitation $\Delta S = 1$ delivers the values of $g$-factors $g^\perp_{1}=3.19$, $g^\perp_{2}=4.76$, and $g^\perp_{3}=7.32$ for $h^\omega \perp B$ as presented by solid lines in Fig.~\ref{fig4:contour}(a), while for $h^\omega \parallel B$ the solid lines in Fig.~\ref{fig4:contour}(b) correspond to $g^\parallel_{1}=5.22$, $g^\parallel_{2}=5.10$, and $g^\parallel_{3}=8.40$.

The linear dependence of the lowest lying $P_1$ mode indicates a gap opening in the field induced paramagnetic phase above $B_{c2}=10$~T.
This is particularly evident for the longitudinal polarization, i.e. $h^\omega \parallel B$, where the lowest-lying $P_1$ starts to harden above $B_{c2}$ [see Fig.~\ref{fig4:contour}(b)].
The clear deviation of the $g_{1}$ and $g_{2}$ factors from the spin-only value indicates the importance of spin-orbit coupling in the system. 
The difference of the \textit{g}-values for the longitudinal and transverse configurations reflects the anisotropy due to the mixing of the higher-lying crystal-field levels of $J_{\text{eff}}=3/2$ and $J_{\text{eff}}=5/2$ with the low-lying one of $J_{\text{eff}}=1/2$, which is also a result of spin-orbit coupling \cite{AbraBlea1970}.
The determined $g_{1}$ and $g_{2}$ values in Na$_2$Co$_2$TeO$_6$ are typical for electron spin resonance of Co$^{2+}$ ions in an octahedral crystal field \cite{AbraBlea1970}.
Therefore, the $P_1$ and $P_2$ can be assigned as single-magnon excitations of the high-field phase. 
In contrast, $g_3$ is about twice large as the values of $g_1$ and $g_2$, 
which is a reminiscence of two-magnon-type excitations as observed in RuCl$_3$ \cite{WangPRLMagn,Ponomaryov20,Sahasrabudhe20,Wulferding2020} and in BaCo$_2$(AsO$_4$)$_2$ \cite{Shi21,Zhang2023}.
 
By changing the external magnetic field be to perpendicular to a honeycomb bond direction, i.e. $B \parallel b$ (see Fig.~\ref{fig1:zero}), we comprehensively characterize field-dependent evolution of the absorption spectra again for two orthogonal polarizations $h^\omega \parallel b$ and $h^\omega \perp b$, the results of which are summarized in Fig.~\ref{fig4:contour}(c)(d).
We note here that the \textit{b} and \textit{a} axes are equivalent with respect to the hexagonal crystal symmetry.
The overall characteristics of the spin dynamics are very similar as for the other field orientation $B \perp a$.
In particular, in the field range between $B_{c1}$ and $B_{c2}$ we observe again a continuum but now for the polarization $h^\omega \perp b$, whereas a sharp absorption profile is observed for $h^\omega \parallel b$.
By comparing these results for the two different field orientations [c.f. Fig.~\ref{fig4:contour}(a)(b) and \ref{fig4:contour}(c)(d)], one can find that the existence of the continuum is not a unique feature for a specific orientation of the applied field, but rather associated with the transverse dynamical spin correlations, i.e. $h^\omega \perp B$. 
In contrast, the dynamic response of the sharp excitation feature is probed exclusively in the longitudinal polarization channel $h^\omega \parallel B$.
A primary difference for the two field orientations appears on the two-magnon $P_3$ mode.
In comparison with $B \perp a$ (i.e. the field along the honeycomb bond direction), for the field perpendicular to the bond direction ($B \parallel b$) the $P_3$ mode is either absent [c.f. Fig.~\ref{fig4:contour}(c) and \ref{fig4:contour}(a)] or observed at lower energies [c.f. Fig.~\ref{fig4:contour}(d) and \ref{fig4:contour}(b)].
These results also indicate that even in the high-field phase a description of the observed quantum spin dynamics in term of free spin waves is not sufficient, but a proper many-body simulation is required (see e.g. Ref.~\cite{Winter2017,Hickey2019,Sahasrabudhe20}.)

To summarize, by performing high resolution terahertz spectroscopy of the quasi-two-dimensional spin-1/2 honeycomb magnet Na$_2$Co$_2$TeO$_6$ in high magnetic fields up to 17~T, we reveal three field regimes which are featured by distinctly different field-dependent spin dynamics and separated by two critical fields around $B_{c1}=7$~T and $B_{c2}=10$~T, respectively.
While below $B_{c1}$ and above $B_{c2}$ the dynamics is characterized by well-defined magnetic excitations, in the intermediate regime $B_{c1}\leq B \leq B_{c2}$ sharp absorption profile and extended continuum are observed in the longitudinal and transverse polarization channels $h^\omega \parallel B$ and $h^\omega \perp B$, respectively, both for the applied fields parallel and perpendicular to the honeycomb bond direction.
The systematic spectroscopic characterization of the spin dynamics points to the importance of Kitaev type interaction in Na$_2$Co$_2$TeO$_6$, and provides a benchmark for a quantitative theoretical description of the magnetic properties. Our results motivate further investigation of the characteristic spin dynamics, e.g. in the full Brillouin zone by inelastic neutron scattering or Raman spectroscopic measurements in high magnetic fields.

\textit{Note added:} Recently, we became aware of related work
on sodium-occupation disorder enriched magnetic excitations~\cite{xiang2023disorderenriched}.

\begin{acknowledgments}
We thank Christian Hess, Lukas Janssen, Huimei Liu, Yuan Wan, and Wang Yang for stimulating discussions. We acknowledge support by the European Research Council (ERC) under the Horizon 2020 research and innovation programme, Grant Agreement No. 950560 (DynaQuanta), by the Estonian Ministry of Education, personal research funding PRG736, and the European Regional Development Fund project TK134.

P.P. and L.P. contributed equally to this work.
\end{acknowledgments}

\bibliographystyle{apsrev4-2}
\bibliography{NCTO_bib}

\end{document}